\documentstyle[aps,pre,multicol,epsf]{revtex}
\begin{document}
\draft
\title{1/f Noise and Extreme Value Statistics}
\vspace {1truecm}
\author{T. Antal${}^1$, M. Droz${}^1$, G. Gy\"orgyi${}^{2}$, and Z. R\'acz${}^{2,3}$}

\address{${}^1$ D\'epartement de Physique Th\'eorique, Universit\'e de
Gen\`eve, CH 1211 Gen\`eve 4, Switzerland}

\address{${}^2$Institute for Theoretical Physics,
E\"otv\"os University, 1117 Budapest, P\'azm\'any s\'et\'any 1/a, Hungary}

\address{${}^3$Laboratoire de Physique Th\'eorique, B\^atiment 210,
Universit\'e de Paris-Sud, 91405 Orsay Cedex, France}

\date{May 31, 2000}

\maketitle

\begin{abstract}
  We study the finite-size scaling of the roughness  
  of signals in systems displaying Gaussian $1/f$ power spectra.
  It is found that one of the extreme value
  distributions (Gumbel distribution) emerges as 
  the scaling function when the boundary conditions are periodic.  
  We provide a realistic example of periodic $1/f$ noise, and 
  demonstrate by simulations that the Gumbel distribution 
  is a good approximation for the case of nonperiodic boundary 
  conditions as well. Experiments on voltage fluctuations 
  in $GaAs$ films are analyzed and excellent agreement is 
  found with the theory. 
\end{abstract}

\date{\today}
\begin{multicols}{2}
\narrowtext


It was about seventy years ago that a current-carrying 
resistor was first observed to exhibit voltage fluctuations with a
power spectra nearly proportional to the inverse of the frequency
\cite{Weissman}. Since then it has been shown that
such an $1/f$ noise is
present in an extraordinary variety of phenomena \cite{hong-kong}, 
examples being  
the light emission of white-dwarfs \cite{press}, the flow of sand
through hourglass \cite{shick},  the ionic current fluctuations in
membranes\cite{siwy}, and the number of stocks traded daily 
\cite{mantegna}.  Most of $1/f$ fluctuations
encountered are Gaussian, although non-Gaussian cases are
known \cite{hong-kong,siwy}. The universality of this scale invariant
phenomenon led to suggestions that part of the explanations 
should come from a generic underlying mechanism. 
Despite a large body of works, however, such a
mechanism has not been discovered yet. 

In this letter, we establish a connection between $1/f$ noise and 
extreme statistics that may provide a new angle at the
generic aspect of the phenomena.
Namely, we shall show that Gaussian $1/f$ power spectra in 
periodic systems imply that the distribution of the 
fluctuations in the finite-size "width" of the signal is one of
the extreme value distributions, the Gumbel distribution \cite{Gumbel}.

As we shall see, the Gumbel distribution emerges as a finite-size scaling 
function in the above connection. Thus the result 
can also be viewed as an interesting contribution to the gallery of 
nonequilibrium scaling functions that
can be effectively used in investigating 
far from equilibrium processes. Indeed,
imagine that a distribution function is
measured in experiments (or simulations). Comparing this function with those
in the theoretically built gallery (note the absence of fitting parameters) 
one can identify the relevant features of the underlying dynamics 
in the experimental (or model) system. 
Such an approach was initiated in connection with surface
growth problems \cite{{FORWZ},{PRZ94},{RP94},{AR96}}, and the results
have been used to establish universality classes in rather diverse
processes such as massively parallel algorithms \cite{Korniss} and the
interface dynamics in the $d=2$ Fisher equation \cite{Tripathy}.  
This line of reasoning \cite{BHF98} has also 
led to a connection between the dissipation fluctuations in a
turbulence experiment \cite{Pinton1-2} and the interface fluctuations
in the $d=2$ Edwards-Wilkinson model ($XY$ model)
\cite{{Holdsworth1-2},{Holdsworth3},{Goldenfeld}} and, 
furthermore, it helped in a parameter-free analysis 
of the upper critical dimension of the Kardar-Parisi-Zhang 
equation \cite{MPPR}. 


The derivation of the Gumbel distribution follows the steps of a similar
calculation for the width distribution of random-walk interfaces
\cite{FORWZ}.  Let the time evolution of the physical quantity of
interest in the interval $0\le t\le T$ be given by $h(t)$.  
This time series
is equivalent to a surface configuration with $h(t)$ being the height
of the surface over a $d=1$ dimensional 
substrate of length $T$ with $t$ being the
coordinate along the substrate.  The quantity of interest is the
mean-square fluctuations of the surface (also called roughness or 
width-square of the surface) given by
\begin{equation}
w_2(h)=\overline{{[\, h(t)-{\overline{h}}\, ]}^2} \, ,
\label{w2}
\end{equation}
where over-bar denotes average of a function over $t$, 
\begin{equation}
\overline F =\frac{1}{T}\int_0^T F(t)dt. 
\end{equation}
Let us assume now that the path probability of a given 
time series $h(t)$ is known  
${\cal P}[h(t)]\sim \exp[{-S[h]}]$.
Then the probability distribution of the surface fluctuations,
$P(w_2)$, can be expressed 
as a path integral \cite{Feynm}
\begin{equation}
P(w_2)  =  \int {\cal D}h(t) \,
\delta\left( w_2 - [\, \overline{h^{\, 2}} \,-\, \overline{h}^{\, 2} \, ] 
\right) 
{\cal P}[h(t)] \ .
\label{Pw_2}
\end{equation}
We shall restrict the above functional integral to periodic
paths $h(t)=h(t+T)$ and, in order to keep ${\cal P}$ normalizable, 
the integration is carried out with $\bar h$ kept fixed.

The next step is to introduce the 
generating function for the moments of $P(w_2)$:
\begin{equation}
G(s)  = \int_{0}^{\infty} dy \,P(y ) e^{-sy} \quad .
\label{G} 
\end{equation}
Substituting  ${\cal P}[h(t)]\sim \exp[{-S[h]}]$ into (\ref{Pw_2}) 
and evaluating the 
integral (\ref{G}), we find the following functional integral
\begin{equation}
G(s) =  {\cal N} \int {\cal D}h(t) \,
\exp\left[- \,S[h] \, - s \,
(\, \overline{h^{\, 2}} \,-\, \overline{h}^{\, 2} \, ) \right] \ ,
\label{Gs}
\end{equation}
where $\cal N$ is a normalization constant to ensure 
$G(0)=1$. 

The key question now is how to choose $S$ in the probability density 
functional for $1/f$ noise.  
The mathematical representation of $1/f$ noise
has been pioneered by Mandelbrot and Van Ness \cite{MandelNess}, 
and since then
a quite intricate theory (involving e.g. fractional 
derivatives) has emerged \cite{Klafter}.  Due to the periodicity imposed
on h(t), however, a simple form for $S$ can be
given in our case.  Namely, we shall consider a "perfect" Gaussian 
$1/f$ noise with the spectrum being linear for all frequencies,
i.e. the following action is assumed
\begin{equation}
S=\sigma \sum_{n=-L}^{L} \vert n \vert \vert c_n\vert^2 
=2\sigma \sum_{n=1}^{L} n \vert c_n\vert^2\, ,
\label{effham}
\end{equation}
where $\sigma$ is a parameter setting the effective surface tension
and the $c_n$-s are the
Fourier coefficients of the signal 
\begin{equation}
h(t) = \sum_{n=-(N-1)/2}^{(N-1)/2} c_{n} e^{2 \pi i n t/T}\,,\hspace{5mm} 
c_{-n} = c_{n}^{\star} \ . 
\end{equation}
Here h(t) is given on $N$ equidistant points ($t=k\Delta t$,
$T=N\Delta t$), and we introduced the notation $L=(N-1)/2$ with 
$N$ assumed to be odd.

Using (\ref{effham}) the functional integral (\ref{Gs}) 
can be written as
\begin{equation}
G(s) =  {\bar{\cal N}} \int {\cal D}[c] \,
\exp\left[- \sum_{n=1}^{L}
2(\sigma  n + s) \vert c_n\vert^2\right ] \, .
\label{Gs2}
\end{equation} 
The integrals over the real and imaginary parts of
$c_{n}$ ($n = 1, 2, \ldots L$) yield 
a simple form for the generating function once we have used the condition $G(0)=1$ 
to determine the normalization
constant $\tilde {\cal N}$ in (\ref{Gs2}):
\begin{equation}
G(s) = \prod_{n=1}^{L} \left(1 + \frac{s}{\sigma n}\right)^{-1}\ .
\label{Gsprod2}
\end{equation}

\begin{figure}[htb]
  \centerline{ \epsfxsize=8cm \epsfbox{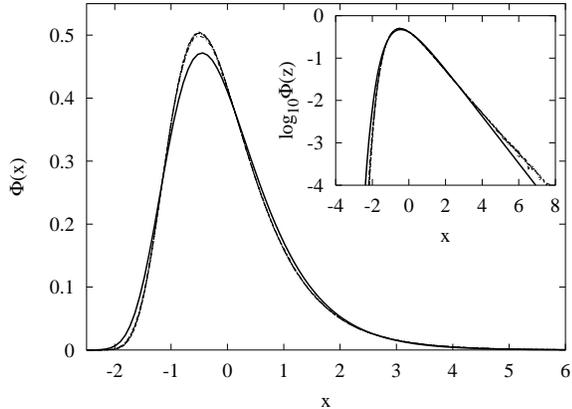} } \vspace{0.5cm}
\caption{The analytical form (\ref{Gumbeldist}) of the roughness  
  distributions for a $1/f$ signal in the periodic case ($N=\infty$,
  solid line) as compared to the numerical result for the
  nonperiodic case for several values of $N$ varying from 64 to 32768
  (dashed lines). Inset shows the same curves on semilog scale.}
\label{fig:nonper}
\end{figure}

The moments of $P(w_2)$ can now be calculated and,
in particular, one finds that the average of $w_2$ diverges 
for large $L$ as
\begin{equation}
\langle w_2 \rangle = \left. - \frac{dG}{ds} \right|_{s=0}
= \frac{1}{\sigma} \sum_{n=1}^{L} n^{-1} \approx \frac{1}{\sigma}
\left[\, \ln L + \gamma \, \right]
\label{wsq}
\end{equation} 
with $\gamma=0.577...$ being the Euler constant. On the other hand, 
the fluctuations of $w_2$ are finite
\begin{equation}
\langle w_2^2 \rangle -\langle w_2 \rangle^2 
= \frac{a^2}{\sigma^2} \,\, ; \hspace{0.5truecm} a=\frac{\pi}{\sqrt{6}} \, .
\label{wssq}
\end{equation}
This means that the scaling variable $w_2/\langle w_2 \rangle$ 
usually considered \cite{FORWZ} in cases when 
$\sqrt{\langle w_2^2 \rangle-\langle w_2\rangle^2}\sim \langle w_2 \rangle$
is not the best choice since the distribution function 
reduces to a delta function. Just as in case of the fluctuations of the $d=2$
EW interface \cite{Holdsworth1-2}, the nontrivial shape underlying the 
delta function can be made visible by introducing
\begin{equation}
x= \frac{w_2-\langle w_2 \rangle}{\sqrt{\langle w_2^2 \rangle-\langle w_2
\rangle^2}} \, . 
\label{scalvar}
\end{equation}
Then the $L\rightarrow \infty$ limit of
the inverse Laplace transform of $G(s)$ yields $P(w_2)$ 
in the following scaling form 
\begin{equation}
\Phi(x)\equiv\sqrt{\langle w_2^2 \rangle -\langle w_2 \rangle^2}P(w_2) = 
\int\limits_{-i\infty}^{i\infty} \frac{ds}{2 \pi i} 
\,e^{x s} \prod_{n=1}^{\infty} 
\frac{ e^{\frac{s}{a n}}}{1+{\frac{s}{a n}}}
\label{Pwsc} \, .
\end{equation}
The infinite product in (\ref{Pwsc}) is equal to 
$e^{(\gamma s/a)}\Gamma(1+s/a)$, so the inverse Laplace transform  
can be evaluated using  
Euler's integral formula for the $\Gamma$ function, and one obtains  
\begin{equation}
  \Phi(x)=ae^{-(ax+\gamma) -e^{-(ax+\gamma)}} \, .
\label{Gumbeldist}
\end{equation}
This scaling function, shown in Fig.~\ref{fig:nonper}, is one of the
central results of our paper. In $\Phi(x)$ one recognizes the Gumbel
distribution \cite{Gumbel} which is one of the three limiting forms of
extreme value statistics.

Extreme statistics has been 
studied in many contexts and the Gumbel distribution emerges 
frequently \cite{Galambos}.
A recent example in connection with surfaces is the study of the 
scaling behavior 
of the growth of the maximal relative height of a surface \cite{shapir}.
Since in most of the these studies an extreme property is investigated, 
it is not entirely surprising to see the Gumbel distribution appearing.
For $1/f$ noise, however, this is not the case.   
A simple quantity such as the mean-square fluctuations 
(roughness of the interface)
is distributed according to extreme (Gumbel) statistics. Although   
we do not see a simple physical reason that necessitates 
this mathematical result, 
we speculate that 
it may be a key feature that underlies a unified treatment of systems 
displaying $1/f$ noise.


When trying to compare the scaling function $\Phi(x)$ with experimental results 
the following problem arises. An experimental signal is analized 
by moving a window of length $L$ and building a histogram 
from the values of $w_2$ computed for the windows. The problem now is that
the boundary conditions for the windows are not periodic and it is known that
the boundary conditions affect the scaling functions \cite{Wu}. Thus 
two questions should be answered. First, is there a physical system 
with an effective action (\ref{effham}) of the $1/f$ noise where 
periodic boundary conditions are realized? 
Second, how sensitive is $\Phi(x)$ to the boundary conditions.

Let us begin with the first question by showing an example of such a system.
We consider the steady state fluctuations 
of a $d=2$ Edwards-Wilkinson (EW) surface with 
the substrate taken to be an infinite plane. We draw a circle of
radius R on the substrate and compute the 
probability density functional of a
height configuration $h_0(\varphi)$ 
over this circle (parametrized by $0\le\varphi<2\pi$). 
We shall find that the action in this functional is equal to 
that of the Gaussian $1/f$ noise given by (\ref{effham}).

In the $d=2$ EW model, the steady state fluctuations of the height 
of a surface $h(\vec x)$
are characterized by the following free-field action
\begin{equation}
  \label{eq:EW-action}
  S_2[h({\vec x})] = \frac{\sigma}{2} \int |\nabla
  h({\vec x})|^2 d^2{\vec x},
\end{equation}
where the integral extends over a plane and the boundary conditions at 
infinity are fixed at $h(\infty)=0$.
Working with polar coordinates $(r,\varphi)$ and specifying the height
function on the $r=R$ circle by $h(r=R,\varphi)=h_0(\varphi)$,
the probability distribution functional of $h_0(\varphi)$ becomes
\begin{equation}
  \label{eq:PDFal-border}
 \hat{P}[h_0(\varphi)] = Z^{-1}
 \int_{h({\vec x})|_R=h_0(\varphi)} {\mathcal
 D}h({\vec x})\,\,  e^{-S_2[h({\vec x})]}\, .  
\end{equation}
Here the functional integration is carried out with the boundary condition
fixed at $r=R$ while $Z$ is the same
integral without restriction on the circle (thus  
${\hat P}[h_0(\varphi)]$ is normalized to 1).  
Since the action is purely quadratic in the field, we have
\begin{equation}
  \label{eq:PDFal-border2}
 {\hat P}[h_0(\varphi)] = Z^{-1}  e^{-S_2[h_{\rm
 c}({\vec x})]}  \int_{\tilde h|_R\equiv 0} {\mathcal
 D}\tilde h({\vec x})\,\,  e^{-S_2[\tilde h({\vec x})]},  
\end{equation}
where $h_{\rm c}({\vec x})$ is the ``classical''
solution that extremizes the action with fixed boundary conditions, and
the integration variable 
$\tilde h({\vec x})=h({\vec x})-h_{\rm c}({\vec x})$ .  
Since the integral does not depend on the
boundary values, we can absorb it into the normalization coefficient,
so the term of interest is the extremal (classical)
action.  The classical field satisfies the Laplace equation 
with fixed boundary condition, that is we have two Dirichlet problems, 
for both inside ($h_{c}^{-}$) and outside ($h_{c}^{+}$)
of the circle with a common boundary condition at $r=R$:
\begin{equation}
  \label{eq:classical-eq}
  \Delta h_{\rm c}^{\pm} ({\vec x}) = 0, \,\,\,\,\,
  h_{\rm c}^{\pm}|_R=h_0(\varphi) \, .  
\end{equation}
Let the solutions of the above problems be $h_c^\pm(r,\varphi)$. 
Substituting them into the action 
(\ref{eq:EW-action}) and applying Gauss' theorem, we can express   
the classical action $S_2[h_c]$ as functional of 
$h_0(\varphi)$. Denoting this functional by $S_c[h_0]$, we have 
\begin{equation}
  \label{eq:classical-action}
  S_c[h_0] = 
  \frac{\sigma R}{2} \int d\varphi \, h_0 (\varphi)\left[
  \left.\partial_r h_c^{-}(r,\varphi)\right |_R -
  \left.\partial_r h_c^{+}(r,\varphi)\right |_R \right]. 
\end{equation}
The solutions of the Dirichlet problems are given by 
\begin{equation}
  \label{eq:classical-field-disk}
  h_c^\pm(r,\varphi) = \frac{1}{\sqrt{2\pi}} \sum_{n=-\infty}^{\infty}
  \left( \frac{R}{r} \right)^{\pm|n|}\, h_{R,n}\, e^{in\varphi},  
\end{equation} 
where $h_{R,n}$ are the Fourier coefficients of $h_0(\varphi)$
satisfying $ h_{R,n} = h_{R,-n} ^{\ast}$.
Substituting (\ref{eq:classical-field-disk}) into 
(\ref{eq:classical-action}) yields the classical action for the 
height fluctuations on the circle
\begin{equation}
  \label{eq:cl-action-disk}
  S_c[h_0(\varphi )] = 2\sigma
   \sum_{n=1}^{\infty}   n\,  |h_{R,n}|^2 \, . 
\end{equation} 
Comparing (\ref{eq:cl-action-disk}) and (\ref{effham}), one finds
that $S_c[h_0(\varphi )]$ is 
the action of a perfect Gaussian 
$1/f$ noise. Since $h_0(\varphi )$ is a 
periodic function, 
we have thus indeed obtained a physical realization of a 
periodic signal with $1/f$ noise.


Let us now turn to the question of how sensitive $\Phi(x)$ 
to changes in the boundary conditions.
Analytically this turns out to
be a hard problem and only numerical calculations were performed.
First, a periodic series of length $\tilde N$ 
having $1/f$ power spectrum was produced using
an appropriately filtered Gaussian white noise.
Next, the signal was divided into non-overlapping segments of length
$N$ and having determined the $w_2$-s, the histogram of $w_2$ and then
the scaling function was built (using the same
normalization as for the periodic boundary condition case).  
In order to obtain satisfactory precision 
we used $\tilde N=2^{24}$ and averaged
over 200 realizations of the periodic signals. 

For $\tilde N = N$ the
result for the periodic case is recovered while, in the 
$N \ll \tilde N$ limit, one finds that $\Phi(x)$ is independent
of the size of the segments in a wide range of $N$ values. 
The $\Phi(x)$ obtained in this $N \ll \tilde N$ limit 
will be considered as 
the scaling function for nonperiodic (or "experimental") boundary conditions. 
As one can see from Fig.~\ref{fig:nonper}, the distribution of
nonperiodic signals deviates from that of the periodic case (Gumbel). The 
deviations, however, are small and mainly
concentrated around the maximum of the function.

\begin{figure}[htb]
  \centerline{ \epsfysize=6cm \epsfbox{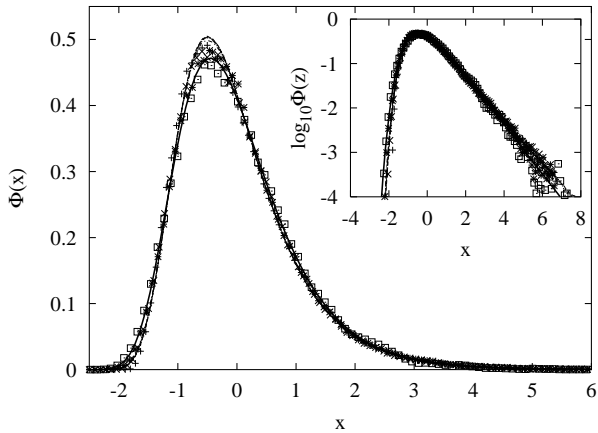} } \vspace{0.5cm}
\caption{Roughness distribution in the experiments 
  on voltage fluctuations
  in resistors calculated for $N=32$ (+), 64 ($\times$), 128 (*), and 256
  ($\Box$), compared to the analytical and numerical results shown in
  Fig.~\ref{fig:nonper}. Inset shows the same curves on semilog scale.}
\label{fig:exp}
\end{figure}

We now consider the case of voltage fluctuations in semiconductor
films.  The experiment was made by A.V. Yakimov and F.N. Hooge
\cite{yaki00}. They considered n-type epitaxial GaAs films grown by
molecular beam epitaxy.  A noise-free current passed longitudinally
through the film and other contacts were used as voltage probes both
for longitudinal and transverse directions.  Several time series with
typically 163840 points were obtained and the power spectrum was found to 
exhibit $1/f$ behavior roughly over two decades\cite{yaki00} of frequencies. 
We have reanalyzed these data by dividing the
signal into segments of length $N = 32$, 64, 128, 256,
and computing the distributions of the roughness for different $N$.  The
results are displayed on Fig.~\ref{fig:exp}. One can see that the
experimental data fit well with the theoretical curves (note that  
no fits are used in collapsing these functions) and that their
precision is not good enough to distinguish between the periodic and
nonperiodic cases.

As a final remark, let us note that 
the apparent ubiquity of the $1/f$ noise is partly the result of
loose terminology. Systems whose power spectrum is of the 
form $1/f^{\alpha}$ with $\alpha$ close to $1$ are also said to exhibit  
$1/f$ noise. The approach we used for the treatment of the 
pure $1/f$ noise can be extended to these systems and the 
dependence of the scaling function $\Phi(x)$ on $\alpha$
can be determined. A detailed discussion of this more general 
problem will be given in a separate paper \cite{papdue}.

We are thankful to A. V. Yakimov for sending us his experimental data,
and to L. B. Kish, Z. Gingl, and L. Sasv\'ari for illuminating 
discussions. This work has been supported by the 
Hungarian Academy of Sciences (Grant No. OTKA T029792) and, partially,
by the Swiss National Science Foundation.



\end{multicols}

\end{document}